\documentclass{article}

\usepackage{arxiv}

\usepackage[utf8]{inputenc} 
\usepackage[T1]{fontenc}    
\usepackage{hyperref}       
\usepackage{url}            
\usepackage{booktabs}       
\usepackage{amsfonts}       
\usepackage{nicefrac}       
\usepackage{microtype}      
\usepackage{cleveref}       
\usepackage{lipsum}         
\usepackage{graphicx}
\usepackage{natbib}
\usepackage{doi}
\usepackage{comment}
\usepackage{algorithmic,eqparbox,array}
\usepackage{algorithm}

\title{Multi-criteria Hardware Trojan Detection: A Reinforcement Learning Approach}

\date{}

\author{ \href{https://orcid.org/my-orcid?orcid=0000-0002-0134-8418}{\includegraphics[scale=0.06]{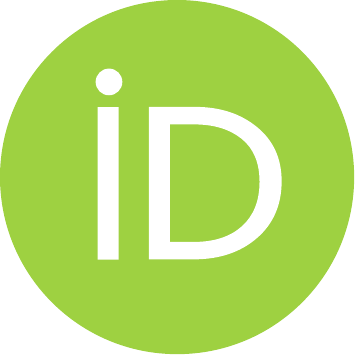}\hspace{1mm}Amin ~Sarihi} \\
	Klipsch School of Electrical\\ and Computer Engineering\\
	New Mexico State University\\
	\texttt{sarihi@nmsu.edu} \\
 	\And
	\href{https://orcid.org/0000-0002-3741-0201}{\includegraphics[scale=0.06]{orcid.pdf}\hspace{1mm}Peter ~Jamieson} \\
	Department of Electrical\\ and Computer Engineering\\
	Miami University\\
	\texttt{jamiespa@miamioh.edu}\\
	\And
 	\href{https://orcid.org/0000-0003-2647-2797}{\includegraphics[scale=0.06]{orcid.pdf}\hspace{1mm}Ahmad ~Patooghy} \\
	Computer Systems Technology\\
	North Carolina A\&T  State University\\
	\texttt{apatooghy@ncat.edu}
 \And
	\href{https://orcid.org/0000-0001-8027-1449}{\includegraphics[scale=0.06]{orcid.pdf}\hspace{1mm}Abdel-Hameed A. ~Badawy} \\
	Klipsch School of Electrical\\ and Computer Engineering\\
	New Mexico State University\\
	\texttt{badawy@nmsu.edu} \\
\\
}

\begin{document}
\maketitle

\begin{abstract}
Hardware Trojans (HTs) are undesired design or manufacturing modifications that can severely alter the security and functionality of digital integrated circuits. HTs can be inserted according to various design criteria, e.g., nets switching activity, observability, controllability, etc. However, to our knowledge, most HT detection methods are only based on a single criterion, i.e., nets switching activity. This paper proposes a multi-criteria reinforcement learning (RL) HT detection tool that features a tunable reward function for different HT detection scenarios. The tool allows for exploring existing detection strategies and can adapt new detection scenarios with minimal effort. We also propose a generic methodology for comparing HT detection methods fairly. Our preliminary results show an average of 84.2\% successful HT detection in ISCAS-85 benchmarks.
\end{abstract}

\keywords{Hardware Trojan, Reinforcement learning, Automated Benchmarks}

\section{Introduction}
Due to time-to-market constraints and increasing production costs, the integrated circuit (IC) supply chain has adopted a multi-party production model. According to this new model, most microelectronic chips are being produced outside of the country~\cite{securing}, raising security concerns about the design and fabrication of chips, particularly hardware Trojan (HT) insertion attacks.

Our current HT detection capabilities suffer from the following shortcomings. 1) Most detection methods perform the HT detection through a one-dimensional lens, i.e., nets' switching activity ~\cite{lyu2020scalable} and~\cite{gohil2022deterrent}. We believe that the current detection methods might not cover the real-world scenarios in which adversaries can insert HTs according to a range of criteria.  2) Available HT benchmarks suffer from significant limitations in size and variety of circuits, as well as the fact that they are all human-crafted and hence are biased by the expert mindset at the creation time ~\cite{sarihi2022hardware}\footnote{The most referenced benchmarks are available on trust-hub.org.}. 

This paper attempts to move the HT detection research space forward by developing a multi-criteria HT detector that explores many HT detection strategies, not limited to a designer's mindset. Our Reinforcement Learning (RL) HT detector has a tunable rewarding function that helps detect different HTs with different insertion strategies. The RL agent explores large circuit  designs promptly and generates test vectors to find HTs in digital circuits. Our threat model consists of a security engineer that must verify a manufactured IC's integrity before allowing it to be integrated into a bigger design. The engineer only can rely on the golden netlist to produce test vectors. Our threat model inherently differs from previous works~\cite{lyu2020scalable}, where the design internals are still accessible in the pre-silicon phase. Our generated test patterns are publicly available through this link\footnote{The link removed for blind review.}.
Additionally, this paper introduces a confidence value as a part of a methodology to compare HT detectors fairly. This helps security engineers to decide the merits of HT detectors for specific applications. In summary, the paper's contributions are as follows:

\begin{itemize}
    \item We introduce an RL-based HT detection tool with a tunable rewarding function that can be modified and re-trained based on different criteria.

    \item We introduce and use a generic methodology to make fair comparisons among HT detectors.
\end{itemize}

The rest of the paper is organized as follows.
We discuss previous endeavors in HT detection in Section~\ref{sec:background}. Section~\ref{sec:proposed} presents our HT detection tool. We define a security metric to better compare the HT detectors by security engineers in Section~\ref{sec:metric}.  Experimental evaluation of the proposed tool and analysis of the results are in Section~\ref{sec:results}. Finally, Section~\ref{sec:conclusion} concludes the paper.

\section{Background and Previous Work}
\label{sec:background}

This section reviews existing hardware Trojan (HT) detection methods. MERO is a test pattern generator that tries to trigger possible HTs by exciting rare-active nets multiple times. MERO becomes less effective with larger circuits. Hasegawa~et al.~\cite{hasegawa2017trojan} extract 51 features from the Trusthub benchmarks and train a Random Forest classifier. However, the studied dataset is limited. Lyu~et al.~\cite{lyu2020scalable} proposed TARMAC, which maps the trigger activation problem to the clique cover problem. TARMAC requires access to the internal nets and testing each suspect circuit separately. TGRL is an RL framework where the agent decides whether to flip a bit in the test vector according to an observed probability distribution. The reward function combines the number of activated nets and their SCOAP~\cite{1585245} (Sandia Controllability/Observability Analysis Program) parameters. The algorithm was not tested on any HT benchmarks. DETERRENT~\cite{gohil2022deterrent} is another RL-based detector that finds the smallest set of test vectors to activate as many rare nets as possible; however, it only targets the switching activity of nets. HW2VEC ~\cite{yu2021hw2vec} uses Graph Neural Networks to extract structural features from graphs and produce graph embeddings. The embeddings are passed to a deep neural  network to classify circuits as HT-free or HT-infected. The detector is trained on Trusthub benchmarks. Unlike the previous work, our study proposes a multi-criteria RL-based HT detector tool that can detect HTs with different insertion strategies.

\section{RL-based HT Detection}
\label{sec:proposed}

From an RL agent perspective for HT detection, the environment is a given circuit (or netlist) to determine whether it is clean or HT-infected. The agent interacts (performs an action) with the circuit by flipping input values to activate internal nets. The RL agent has an $n$-dimensional binary action space $a_t=[a_1,a_2,..., a_n]$ where $n$ is the number of circuit primary inputs. The agent may set or reset each $a_i$ to transition to another state. $a_i=0$ denotes that the value of the $i^{th}$ input will remain unchanged from the previous test pattern, and $a_i=1$ means that the input bit will flip. Attackers are likely to choose trigger nets with a consistent value (either $0$ or $1$) most of the time. Thus, a detector aims to activate as many dormant nets as possible. We consider two different approaches for identifying such rare nets:

\textbf{1) Dynamic Simulation}: We feed each circuit with $100K$ random test patterns and record the value of each net. Through logging nets transitions, we populate the switching activity statistics for each net and compare it against a threshold $\theta$ (ranges in $[0,1]$). Nets with switching below $\theta$ are considered rare nets.

\textbf{2) Controllability Simulation}: This approach classifies the nets based on their \textit{controllability\footnote{Controllability is the difficulty of setting a particular net to $0$ or $1$ logic value.}} values. Low switching nets have a high difference between their controllability value~\cite{sebt2018circuit}, \textit{i.e.}, they are mostly stuck at $0$ or $1$. We set a threshold value $\eta$ as defined in Eq.~\ref{eq1}:
\begin{equation}
    \eta=\frac{|CC1(Net_i)-CC0(Net_i)|}{Max(CC1(Net_i),CC0(Net_i))}
    \label{eq1}
\end{equation}
where $CC0(Net_i)$ and $CC1(Net_i)$ are the combinational controllability of $0$ and $1$ for $Net_i$, respectively. The $\eta$ parameter ranges between $[0,1)$ such that higher values of $\eta$ correlate with lower net activity~\cite{sebt2018circuit}. 

\begin{figure*}[!t]
  \centering
  \includegraphics[width=1\textwidth]{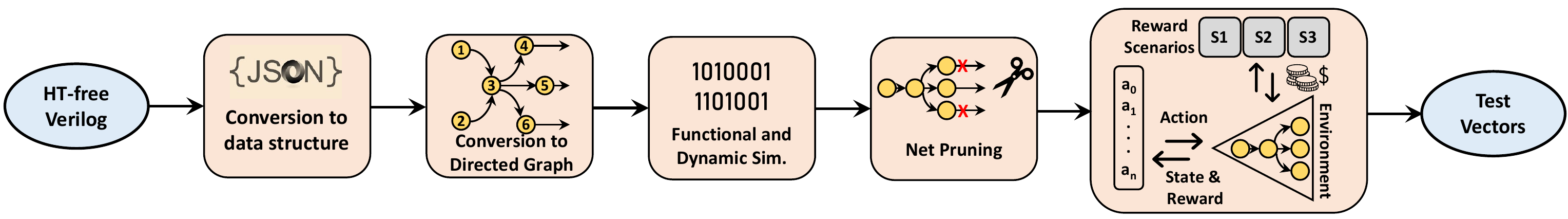}
  \caption{The proposed toolset workflow.}
  \label{flow}
\end{figure*}

Our RL state is mapped to the set of the collected rare nets. In a circuit with $m$ rare nets, the state space is defined as $State_t=[s_1, s_2, ..., s_m ]$ where $s_i$ is associated with the $i^{th}$ net in the set. Whenever an action (a test pattern) activates $s_i$ (taking its rare value), it will set that state to $1$ in the state vector. Otherwise, its state stays at $0$. As can be inferred, the action and state spaces are multi-binary. Figure~\ref{flow} summarizes our tool flow. 

\subsection{Rewarding Functions}

The agent's goal is to activate as many HT triggers as possible. Thus, a part of the rewarding function should enumerate rare nets. However, we should avoid over-counting situations where a rare net has successive dependent rare nets. We adopt a pruning strategy and pick the rarest net in a sequence of dependent rare nets (seen in Figure 1).

This policy will help accelerate the RL agent to converge on the global minima faster.

As for rewarding the agent, we consider three rewarding functions, and we explain them in the rest of this section. In our first rewarding function (hereafter $D1$), we use a copy of the agent's previous state and encourage it to generate states that differ from the previous one. This pushes the agent towards finding test vectors that lead to unseen states. The pruned current and previous state vectors are passed as inputs to $D1$; the final reward is the output. The reward function comprises an \textit{immediate} and \textit{sequential} parts. The sequential reward is computed by making a one-to-one comparison between the nets in the old and new states. The highest reward is given when an action can activate a net that was not triggered in the previous state, where it is given $+40$ for each net. If a rare net continues to be active in the new state, the agent will still be rewarded $+20$. The worst state transition is whenever an agent takes an action that leads to a rare net losing its rare value, and that is rewarded $-3$. Lastly, if the agent cannot activate a rare net after a state transition, it will be rewarded $-1$. The immediate award is the number of activated rare nets in the new state. Lastly, the final reward is a weighted mixture of immediate and sequential rewards with tunable weights.

\begin{algorithm}[!t]
    \caption{Rewarding Function $D2$}
    \begin{flushleft}
    \hspace*{\algorithmicindent}\textbf{\textit{Input: }}{Net switching vector} $Switching_{vector}$,\\ \hspace*{\algorithmicindent}\textbf{}{Current state vector $State_{vector}$, State Vector Length $K$ }\\
    \hspace*{\algorithmicindent}\textbf{\textit{Output: }}{Final reward $Reward_{final}$}\\
    \end{flushleft}
     \begin{algorithmic}[1]
        \STATE $Reward_{vector} = [0] * K$ 
        \FOR{$k \in \{0,\dots,K-1\}$}
            \IF {($Switching_{vector}[k] != 0$)}
                \STATE $Reward_{vector}[k] = Switching_{vector}[k]^ {-1} $
            \ELSE
                \STATE $Reward_{vector}[k]= 0$
            \ENDIF
        \ENDFOR
        \STATE $reward_{max} = max(Reward_{vector}[~]$)
        \FOR{$k \in \{0,\dots,K-1\}$}
            \IF {($Switching_{vector}[k] == 0$)}
                \STATE $Reward_{vector}[k] = 10 * reward_{max}  $
            \ENDIF
        \ENDFOR
        \STATE $Reward_{final} = 0$
        \FOR{$k \in \{0,\dots,K-1\}$}
            \IF {($State_{vector}[k]==1$)}
                \STATE $Reward_{final} += Reward_{vector}[k]$
            \ELSE
                \STATE  $Reward_{final} += -1$
            \ENDIF
        \ENDFOR
        
    \end{algorithmic}
    \label{alg2}    
\end{algorithm}

Algorithm~\ref{alg2} describes our second rewarding function $D2$. In this case, the agent gains a reward proportional to the difficulty of the rare net it can trigger. This reward is computed using the inverse of net switching activities (line $4$). If no vectors were found to trigger a net, it would be rewarded $10X$, the greatest reward in the vector (line $12$). The algorithm encourages the agent to trigger the rarest nets in the circuit.

In the third rewarding function ($D3$), rare nets are populated based on threshold $\eta$ in Eq.~\ref{eq1}. When a rare net is activated, the agent is rewarded with the controllability of the rare value. This scenario aims to investigate controllability-based HT detection using an RL algorithm.

\section{The Proposed Generic HT-Detection Metric}
\label{sec:metric}

We propose the following methodology to the community for fair and repeatable comparisons among HT detection methods. This methodology obtains a confidence value that one can use to conduct a fair comparison between different HT detection methods. There are $4$ possible outcomes when an HT detection tool studies a given circuit. From the tool user's point of view, the outcomes are probabilistic events. For example, when an HT-free circuit is being tested, the detecting tool may either classify it as an infected or a clean circuit, \textit{i.e.}, $Prob(FP) + Prob(TN) = 1$ where $FP$ and $TN$ stand for \textit{False Positive} and \textit{True Negative} events. Similarly, for HT-infected circuits, we have $Prob(FN) + Prob(TP) =1$. 
We know $FN$ and $FP$ are two undesirable outcomes that detectors misclassify. Between these two, $FN$ cases are much more dangerous because an $FN$ case leads to a situation in which we rely on an HT-infected chip, whereas an $FP$ case means wasting a clean chip by either not selling or not using it. So, we need to know how HT detection tools' user (might be a security engineer or a company representative) prioritizes $FN$ and $FP$ cases. We define a parameter $\alpha$ as the ratio of the undesirability of $FN$ over $FP$. The tool user determines $\alpha$ based on characteristics and details of the application that eventual chips will be employed in, \textit{e.g.}, the risks of using an infected chip in a device with a sensitive application versus using a chip for home appliances. Note that the user sets this value, which is not derived from the actual $FP$ and $FN$. After $\alpha$ is set, it is plugged in Eq.~\ref{equ:metric} and a general confidence basis $ Conf. \, Val $ is computed. 

\begin{equation}
    \label{equ:metric}
    Conf. \, Val = \frac{(1-FP)}{(1/\alpha+FN)}
\end{equation}
This metric can make a fair comparison between HT detection methods regardless of their detection criteria and implementation methodology. The defined confidence metric combines the two undesirable cases concerning their severity from the security engineer's point of view, and it ranges between $[\frac{0.5\alpha}{1+0.5\alpha}..\alpha]$. The closer the value is to $\alpha$, the higher the confidence in the detector. The absolute minimum of the $Conf. \, Val = 1/3$ that happens when $\alpha=1$ and $FP = FN = 50\%$. This analysis assumes that $FN$ and $FP$ are independent probabilities. We note that for some detection methods, $FP$ is always $0$. For instance, test-based HT detection methods that apply a test pattern to excite HTs use a golden model (HT-free) circuit for comparison and decision-making. There is no way for such methods to detect an HT in a clean circuit falsely. However, our metric is general and captures such cases.

\section{Experimental Evaluations}
\label{sec:results}
Our proposed multi-criteria HT detector is developed in Python. The training process of the RL agent is done using the PPO (proximal policy optimization)~\cite{schulman2017proximal} from the Stable Baselines library with an episode length of $10$. This guarantees that the agent would reset each $10$ episodes and agent observes a new state. We select six circuits from ISCAS-85, namely $c432$, $c880$, $c1355$, $c1908$, $c3540$, and $c6288$. 

To accelerate the training of the RL agent, instead of calling time-consuming graph functions, we built adjacency matrices and dictionaries that contain structural information of each node within the graph. This simple yet efficient technique speeds up training and testing processes by $3.7\times$ and $3.2\times$, respectively.
\begin{table*}[!ht]
\caption{Detection accuracy of scenarios D1, D2, and D3 for HTs with different input widths in \cite{sarihi2022hardware}.}
\label{tab3}
\begin{center}
\scalebox{0.92}{
\begin{tabular}{| c | c | c | c | c | c | c | c | c | c | c | c | c | c | c | c }
\hline
\textbf{Benchmark} & \multicolumn{3}{ c |}{\textbf{2-Input HT }} & \multicolumn{3}{ c |}{\textbf{3-Input HT }} &\multicolumn{3}{ c |}{\textbf{4-Input HT }}  & \multicolumn{3}{ c |}{\textbf{5-Input HT }}  \\ 
\cline{2-13}
   & \textbf{  D1} & \textbf{  D2} & \textbf{  D3}    & \textbf{  D1} & \textbf{  D2} & \textbf{  D3}    & \textbf{  D1} & \textbf{  D2} & \textbf{  D3}   & \textbf{  D1} & \textbf{  D2} & \textbf{  D3} \\ \hline
c432   & 15.0\% & 27.8\% & 8.0\% & 41.2\% & 61.9\% & 42.8\% & 36.4\% & 66.5\% & 24.9\% & 24\% & 49.1\% & 21.6\%  \\ \hline
c880   & 100\% & 100\% & 100\% & 100\% & 92.0\% & 84.0\% & 86.7\% & 83.0\% & 64.1\% & 85.1\% & 79.3\% & 40.2\%  \\ \hline
c1355 & 94.6\% & 99.1\% & 98.3\% & 92\% & 98.1\% & 97.5\% & 90.5\% & 97.7\% & 96.2\% & 89.6\% & 97.0\% & 95.5\%  \\ \hline
c1908  & 96.4\% & 98.3\% & 97.3\% & 97.6\% & 96.4\% & 94.9\% & 93.0\% & 93.2\% &94.4\% & 89.6\% & 91.1\% & 86.7\%  \\ \hline
c3540  & 56.0\% & 89.8\% & 89.8\% & 86.5\% & 91.6\% & 91.0\% & 89.4\% & 97.1\% & 98.8\% & 77.8\% & 79.2\% & 83.9\%  \\ \hline
c6288    & 96.1\% & 97.9\% & 97.1\% & 97.5\% & 97.6\% & 97.2\% & 95.6\% & 96.1\% & 95.9\% & 93.3\% & 94.1\% & 93.7\%  \\ \hline
\textbf{Confidence Value}    & \textbf{2.97} & \textbf{4.07} & \textbf{3.19} & \textbf{4.13} & \textbf{4.89} & \textbf{3.93} & \textbf{3.56} & \textbf{4.73} & \textbf{3.23} & \textbf{2.91} & \textbf{3.52} & \textbf{2.51}  \\ \hline
\end{tabular}
}
\end{center}
\end{table*}
We start from $450K$ of timesteps for training in $c432$ and increase the timesteps for each successive circuit by $10\%$ to enable enough exploration for larger circuits. We ran the training processes in parallel for each circuit. This process took nearly $27$ hours to train the benchmark set. In the testing phase, we ran the trained RL agent for $20K$ episodes. To select a test vector, we set a cut-off reward of one-tenth of the collected reward in the last training episode (since we have ten steps per episode). We gather $20K$ test vectors that surpass this reward threshold. 

Table~\ref{tab3} summarizes the detection percentages of our three detection scenarios for different HT sizes inserted in ISCAS-85~\cite{sarihi2022hardware}. The inserted HTs in this dataset were introduced to address two issues: 1) removing inherent human bias in current HT databases and 2) providing ample HT instances for training detectors. Table~\ref{tab3} lists six benchmarks with HTs triggered by $2$, $3$, $4$, and $5$ input wires and reports the detection accuracy for $D1$, $D2$, and $D3$ (labeled across the top of the table). The number of HTs for each case is in~\cite{sarihi2022hardware}.

From the table, $D2$ has the best detection rate in most cases; however, exceptions exist. For instance, in $c880$, the detection rate for $D1$ is equal to or better than $D2$, especially for 5-input HTs. The same happens for 3-input HTS in $c1908$.

On the other hand, $D3$ shows its superiority in $c3540$. Except for the 3-input Trojans, $D3$ equals or is better than the other two rewarding scenarios. This underlines the importance of $D3$, which uses an inherently different detection criterion. Polling among the three HT detection scenarios can generally lead to satisfactory HT detection in most circuits.

One interesting observation concerns the detection rate of $c432$. While applying $100,000$ random test vectors, we found that the rarest net in the circuit was triggered $7\%$ of the time, which is significantly higher than other circuits where many nets exhibit switching activity of less than $1\%$. This suggests that the inserted HTs in the $c432$  might be activated easier with random test vectors. To test this hypothesis, we generated $20,000$ additional random test vectors and applied them to the circuit, detecting $99\%$ of the HTs. This demonstrates that the RL attack did not have the intended impact in $c432$.

The confidence metric of our HT-detection tool proposed in Section~\ref{sec:metric} can be seen in the last row of the table, assuming $\alpha=10$. The confidence value for each column is calculated by averaging each detection scenario for each column. The table shows that the security engineer can put confidence in the $D2$ detector since it has higher confidence values than the other detection scenarios. The confidence value only surpasses $5$ for detectors with $FN < 10\%$. In other words, the confidence value increases sharply, and better HT detectors are rewarded with more exponential confidence values.

\section{Conclusion}
\label{sec:conclusion}
 This paper  emphasizes the need for multi-criteria HT detection tools and universal metrics to compare them. We propose a reinforcement learning tool for hardware Trojan detection, which features three rewarding functions that detect a wide range of HTs. Results on ISCAS-85 circuits showed a high detection rate of the proposed tool for various HTs. We also present a methodology to help the community compare HT detection methods regardless of their implementation details. We applied the methodology to our HT detection and discovered that our tool offers the highest confidence in HT detection when using the rewarding function D2.

\bibliographystyle{ACM-Reference-Format}
\bibliography{references}

\end{document}